\documentstyle[11pt,newpasp,twoside]{article}
\def\edcomment#1{\iffalse\marginpar{\raggedright\sl#1\/}\else\relax\fi}
\reversemarginpar

\begin{document}
\title{Photometric Properties of Low Redshift Galaxy Clusters}
 \author{W.A. Barkhouse and H.K.C. Yee}
\affil{Department of Astronomy, Universiy of Toronto, 60 Saint George St. 
Toronto, Ont. M5S 1A7, Canada}
\author{O. L{\'o}pez-Cruz}
\affil{INAOE-Tonantzintla, Tonantzintla Pue. M{\'e}xico}

\section{Preliminary Results}

A recent comprehensive photometric survey of 45 low-{\it z} X-ray selected
 Abell 
clusters (L{\'o}pez-Cruz 1997) has measured significant variations in the 
faint end slope of the luninosity function (LF). This result has indicated
 that 
dwarf galaxies (dGs) have different mixtures in relation with the cluster
 environment. Clusters having a central ``cD-like'' galaxy have a flatter 
faint end slope than non-cD clusters. Also, cD clusters were found to have 
a dwarf-to-giant ratio (D/G) which was smaller than non-cD clusters. 
L{\'o}pez-Cruz et al. (1997) has suggested that the light contained in cD envelopes 
can be accounted for by assuming that it is produced from stars that 
originally formed dGs. In this simple model, the D/G would be expected to
 increase with radial distance from the cluster centre due to the decrease  
in the disruptive forces. 

In order to test the dG disruption model, {\it B} and {\it R} band images
 of a sample of 
27 low-{\it z} ($0.02 \leq z \leq 0.04$) Abell clusters have been obtained 
with the 8k CCD mosaic camera on the KPNO 0.9m telescope. This
 telescope/detector combination provides a $1^{o}\times 1^{o}$ field of 
view, giving an areal coverage of $1-2h^{-1}$ Mp$c^{2}$.
These observations will allow us to 
probe several magnitudes deeper than the L{\'o}pez-Cruz (1997) survey and 
provide a definitive measure of the dG LF. 
Preliminary LFs 
and D/G ratios have been calculated
 for five clusters (A1185, A1656, A2151, A2152, and A2197). 
A significant increase
 in the faint end slope between the inner (0.0-0.75 Mpc) and outer
(0.75-1.50 Mpc) LF can be seen for A2151
 ($H_{o}=~50~{\rm km~s^{-1}~Mpc^{-1}}$).
 This indicates that the number of dGs, defined as the ratio of the number 
of galaxies with $-19\leq M_{R} \leq -15$ to those with $M_{R} < -19.5$, 
 has increased in the outer radial bin
 as compared to the inner cluster region.
All five clusters also show a 
significant dip in the LF at $M_{R} \sim -19$. This dip suggests that the LF
 can be 
modelled by 2 components: a log-normal bright component, and a Schecheter 
function faint component.

\end{document}